\documentclass[10pt,conference]{IEEEtran}
\IEEEoverridecommandlockouts
\usepackage{amsmath,amssymb,amsfonts}
\usepackage{balance}
\usepackage{graphicx}
\usepackage{xcolor}
\usepackage{booktabs}
\usepackage{subfig}
\usepackage{textcomp}
\usepackage{utfsym}
\usepackage{listings}
\usepackage{algorithm}
\usepackage[noend]{algpseudocode}
\usepackage{amsmath}
\usepackage{array}
\usepackage{multirow}
\usepackage{pifont}
\usepackage{afterpage}
\usepackage[normalem]{ulem}
\usepackage{xspace}
\usepackage{enumitem}
\usepackage[most]{tcolorbox}
\usepackage{hyperref}  
\usepackage{cleveref}  
\usepackage{url}
\usepackage{tabularx}
\usepackage{booktabs} 
\usepackage[table]{xcolor}
\newcolumntype{Y}{>{\centering\arraybackslash}X} 
\newcolumntype{A}{>{\hsize=.3\hsize\centering\arraybackslash}X}  
\newcolumntype{B}{>{\hsize=.5\hsize\centering\arraybackslash}X}  
\def\BibTeX{{\rm B\kern-.05em{\sc i\kern-.025em b}\kern-.08em
    T\kern-.1667em\lower.7ex\hbox{E}\kern-.125emX}}

\tcbset{
  colback=gray!15,    
  colframe=gray!20,   
  arc=1mm,            
  boxrule=0.5pt,      
  width=\linewidth,   
  left=5pt,           
  right=5pt,          
  top=5pt,            
  bottom=5pt          
}

\newcommand{\majorR}[1]{\textcolor{black}{#1}} 

\newcommand{\tool}{\textsc{Peace}\xspace}
\newcommand{\benchmark}{\textsc{PeacExec}\xspace}

\def\BibTeX{{\rm B\kern-.05em{\sc i\kern-.025em b}\kern-.08em
    T\kern-.1667em\lower.7ex\hbox{E}\kern-.125emX}}

\begin{document}
\title{\tool: Towards Efficient Project-Level \majorR{Efficiency} Optimization via Hybrid Code Editing}

\author{%
  \IEEEauthorblockN{%
    Xiaoxue Ren\IEEEauthorrefmark{2},
    Jun Wan\IEEEauthorrefmark{2},
    Yun Peng\IEEEauthorrefmark{3},
    Zhongxin Liu\IEEEauthorrefmark{2}\IEEEauthorrefmark{1},
    Ming Liang\IEEEauthorrefmark{4},
    Dajun Chen\IEEEauthorrefmark{4},
    Wei Jiang\IEEEauthorrefmark{4},
    Yong Li\IEEEauthorrefmark{4}%
  }%
  \IEEEauthorblockA{\IEEEauthorrefmark{2}Hangzhou High-Tech Zone (Binjiang) Institute of Blockchain and Data Security, Zhejiang University, Hangzhou, China\\
  }
  \IEEEauthorblockA{\IEEEauthorrefmark{3}The Chinese University of Hong Kong, Hong Kong, China\\
  }
  \IEEEauthorblockA{\IEEEauthorrefmark{4}Ant Group, Hangzhou, China\\
  \{xxren, 22451014, liu\_zx\}@zju.edu.cn, ypeng@cse.cuhk.edu.hk, 
  \{liangming.liang, chendajun.cdj, jonny.jw, liyong.liy\}@antgroup.com
  }
  \thanks{\IEEEauthorrefmark{1} Corresponding author.}
}



\maketitle

\begin{abstract}
Large Language Models (LLMs) have demonstrated significant capability in code generation, but their potential in \majorR{code efficiency optimization} remains underexplored. Previous LLM-based \majorR{code efficiency optimization} approaches exclusively focus on function-level optimization and overlook interaction between functions, failing to generalize to real-world development scenarios. Code editing techniques show great potential for conducting project-level optimization, yet they face challenges associated with invalid edits and suboptimal internal functions.
To address these gaps, we propose \tool, a novel hybrid framework for \textbf{P}roject-level code \textbf{E}fficiency optimization through \textbf{A}utomatic \textbf{C}ode \textbf{E}diting, which also ensures the overall correctness and integrity of the project.
\tool integrates three key phases: dependency-aware optimizing function sequence construction, valid associated edits identification, and \majorR{efficiency optimization} editing iteration. 
To rigorously evaluate the effectiveness of \tool, we construct \benchmark, the first benchmark comprising 146 real-world optimization tasks from 47 high-impact GitHub Python projects, along with highly qualified test cases and executable environments. Extensive experiments demonstrate \tool's superiority over the state-of-the-art baselines, achieving a 69.2\% correctness rate (pass@1), +46.9\% opt rate, \majorR{and 0.840 speedup} in execution efficiency. 
Notably, our \tool outperforms all baselines by significant margins, particularly in complex optimization tasks with multiple functions. 
Moreover, extensive experiments are also conducted to validate the contributions of each component in \tool, as well as the rationale and effectiveness of our hybrid framework design.
\end{abstract}


\section{Introduction}\label{sec:intro}

Large Language Models (LLMs) have demonstrated remarkable performance in code intelligence tasks, particularly in code generation~\cite{achiam2023gpt, dubey2024llama, guo2024deepseek}. 
Given natural language descriptions, they are capable of generating syntactically correct and functionally meaningful code, thus facilitating automated software development and largely improving productivity. 
This could be demonstrated by the 89\% pass@1 of the state-of-the-art (SOTA) LLM on the EvalPlus benchmark~\cite{Liu2023is}.

Despite LLMs' great success in code generation, code quality assurance, especially for \majorR{code efficiency optimization}, which aims to improve the time efficiency of functionally correct code, is an urgent yet challenging task for LLMs. In real-world software development, code \majorR{efficiency} is a critical factor that impacts system scalability, efficiency, and maintainability~\cite{Garg2022deepdev, zelikman2024self, garg2023rapgen, Peng2025coffe}. 
By analyzing 2,000 top-starred popular open-source Python projects from GitHub, we find that 41.25\% (825 out of 2,000) of the projects have issues explicitly related to \majorR{code efficiency} optimization, explicitly by keywords related to efficiency (i.e., efficiency, speedup, etc.) in their issue reports.
It further demonstrates the urgent demand for automated solutions that assist developers in optimizing the \majorR{code efficiency} of projects.

There are some research efforts~\cite{shypula2023learning, gao2024search, gong2025language, huang2024effi, garg2023rapgen,zelikman2024self, huang2024soap} being devoted to LLM-based \majorR{code efficiency optimization}.
For example, Shypula et al.~\cite{shypula2023learning} propose a broad range of adaptation LLM-based strategies for \majorR{code efficiency optimization}, including retrieval-based few-shot prompting, chain-of-thought, as well as fine-tuning.
Garg et al.~\cite{garg2023rapgen} solve performance bugs by retrieving relevant instructions from a knowledge base of past bug fixes and using them to generate a prompt, which is fed into an LLM in a zero-shot manner.
SBLLM~\cite{gao2024search} formulates code \majorR{efficiency} optimization as a search problem by integrating LLMs with evolutionary methods for iterative refinement.
Huang et al.~\cite{huang2024soap} propose a self-optimization process that leverages open-source LLMs to generate a high-quality dataset, which is then used to fine-tune LLMs for code \majorR{efficiency} optimization.

However, \textbf{a key limitation is that prior work overlooks the complex function interactions when conducting \majorR{code efficiency optimization}, which hinders generalization to real-world scenarios}.
In real-world development, \majorR{code efficiency optimization} often involves understanding project-wide context and modifying multiple functions. To address this, we reformulate project-level \majorR{efficiency} optimization as a code editing task that targets multiple functions, rather than generating or repairing a single one. Prior work~\cite{gupta2023grace, liu2024coedpilot} shows that leveraging associated edits effectively guides such multi-function modifications.

To implement code editing techniques for project-level \majorR{code efficiency optimization}, we need to determine what code to edit, when to edit it, and how to edit it:
\begin{itemize}[leftmargin=*]
    \item \emph{What to edit:} First, it is essential to identify the functions to be edited for \majorR{efficiency} optimization. 
    Although the target function is pre-defined before optimization, its efficiency may be largely affected by other functions due to call dependencies. Therefore, it is usually necessary to edit multiple functions at the same time to improve the \majorR{efficiency} of the target function.
    \item \emph{When to edit:} After we collect the functions to edit, we need to decide their editing order based on the call relationships. As some functions may not have direct call relationships, it is important to edit them in an appropriate order to avoid the change of one function negatively affecting the change of another function.
    \item \emph{How to edit:} With an ordered sequence of functions to edit, we need to know how to edit each function one by one. Previous studies~\cite{gupta2023grace, liu2024coedpilot} reveal that the history edit records in the project provide valuable insights about how the project was previously optimized and could be used to guide future edits. However, as a vast number of edit records accumulate over time in large-scale software projects, \textbf{\emph{invalid associated edits could adversely affect optimization guidance} (C1)}. In addition, project-level information may be misleading in code efficiency optimization, as the implementations of some internal functions in the projects may be suboptimal. This indicates that \textbf{\emph{only considering internal functions in projects may result in suboptimal \majorR{efficiency} optimization} (C2)}.
\end{itemize}
To implement the above workflow and solve the challenges of project-level efficiency optimization, we propose \tool, a novel hybrid framework for \textbf{P}roject-level \majorR{\textbf{E}fficiency} optimization through \textbf{A}utomatic \textbf{C}ode \textbf{E}diting.
Specifically, \tool consists of three phases: 
\textbf{(1) Dependency-aware optimizing function sequence construction}.
It focuses on identifying and ordering the functions to be optimized, addressing both the \textit{what to edit} and \textit{when to edit} problems. By analyzing the relevance between the target function and its caller and callee functions, we construct an optimizing function sequence to ensure \majorR{efficiency} improvements are applied in a consistent and dependency-aware manner.
\textbf{(2) Valid associated edits identification}, which is designed to address \textbf{C1} in \textit{how to edit} problem.
This phase iteratively retrieves and filters historical edits to identify valid associated edits that offer meaningful guidance for optimization. 
By combining dependency analysis with LLM-based semantic assessment, we ensure that only relevant and beneficial historical edits are used to guide the optimization.
\textbf{(3) \majorR{Efficiency optimization} editing iteration}, which is designed to address \textbf{C2} in \textit{how to edit} problem.
This phase iteratively edits the functions in the constructed sequence using a fine-tuned \majorR{efficiency} optimizer. The optimizer leverages both internal and external high-performance implementations to generate more efficient solutions. Through repeated iterations, it enhances the project's overall efficiency while guaranteeing correctness and stability.

Since existing benchmarks rarely focus on project-level efficiency optimizations and cannot evaluate our \tool, we construct a benchmark for the project-level code efficiency optimization task, named \benchmark.
\benchmark contains 146 optimization tasks collected from 47 popular Python GitHub projects, covering 80 single-function and 66 multi-function optimization tasks.
Each optimization task in our \benchmark consists of a target function for optimization, the corresponding executable project, a task prompt, the historical edits, and the test cases for evaluation.
Note, the task prompt is extracted from \majorR{efficiency optimization} related commit messages. If no explicit task prompt exists, we use general \majorR{efficiency} optimization instructions as the default input.
The historical edits are from previous commits, and the test cases are collected from the efficiency-related commits.

We evaluate the performance of \tool with \benchmark regarding both editing correctness and execution efficiency improvement by measuring pass@1, opt rate and speedup, respectively. 
Unlike existing function-level code
optimization techniques, where the opt rate is measured by comparing the execution performance before and after optimization,
our project-level tasks sourced from GitHub lack pre-optimization test cases. Inspired by prompt-based optimization techniques~\cite{zelikman2024selftaught, shypula2023learning}, we leverage the strong optimization capabilities of LLMs (e.g., GPT-3.5, GPT-4o) and measure the opt rate by comparing the results to those produced by GPT-4o with instruction prompting. 
\majorR{Then, we also compare the efficiency of various optimization methods with the ground truth (e.g., the original human-optimized patches) of our collected tasks.}
Note, following Peng et al.~\cite{Peng2025coffe}, we adopt CPU instruction count as a stable and consistent metric for evaluating code efficiency, offering a more reliable alternative to execution time.

Extensive experiments illustrate that our \tool achieves 69.2\% of pass@1 and +46.9\% of opt rate compared to the GPT-4o instruction-prompting method, which outperforms all selected baselines (e.g., CoEdPilot~\cite{liu2024coedpilot}, SBLLM~\cite{gao2024search}, and DeepDev-PERF~\cite{Garg2022deepdev}). 
\majorR{Additionally, our \tool achieves 0.840 speedup, which indicates
that \tool's optimization performance is closest to that
of human-written patches, outperforming all other baselines.}
Moreover, we conduct ablation studies to illustrate the effectiveness of different components of our \tool.
We then experiment to investigate the performance of the fine-tuning performance optimizer in our \tool, which can explain the rationality of the hybrid design of our framework.
We also validate that our \tool is well-designed and can excel in both single-function and multi-function optimization tasks, demonstrating robustness across varying task complexities.


In summary, our contributions are as follows:
\begin{itemize}[leftmargin=*]
    \item We propose \tool, a novel hybrid framework for project-level \majorR{efficiency} optimization via automatic code editing, ensuring the overall correctness and integrity of the project.
    \item We construct the first benchmark for project-level code efficiency optimization tasks (i.e., \benchmark), which contains 146 optimization tasks from 47 popular Python GitHub projects, covering 80 single-function and 66 multi-function efficiency optimization tasks.
    \item Extensive experiments compare \tool with SOTA baselines to evaluate correctness and efficiency on project-level optimization tasks. Furthermore, we validate the rationale and effectiveness of our hybrid framework design through additional experiments.

    \item We open-source the replication package~\cite{ourpackage}, including \benchmark, the source code of \tool, and results.
\end{itemize}





\section{Motivation}

\begin{figure}[ht]
    \centering
    \includegraphics[width=0.9\linewidth]{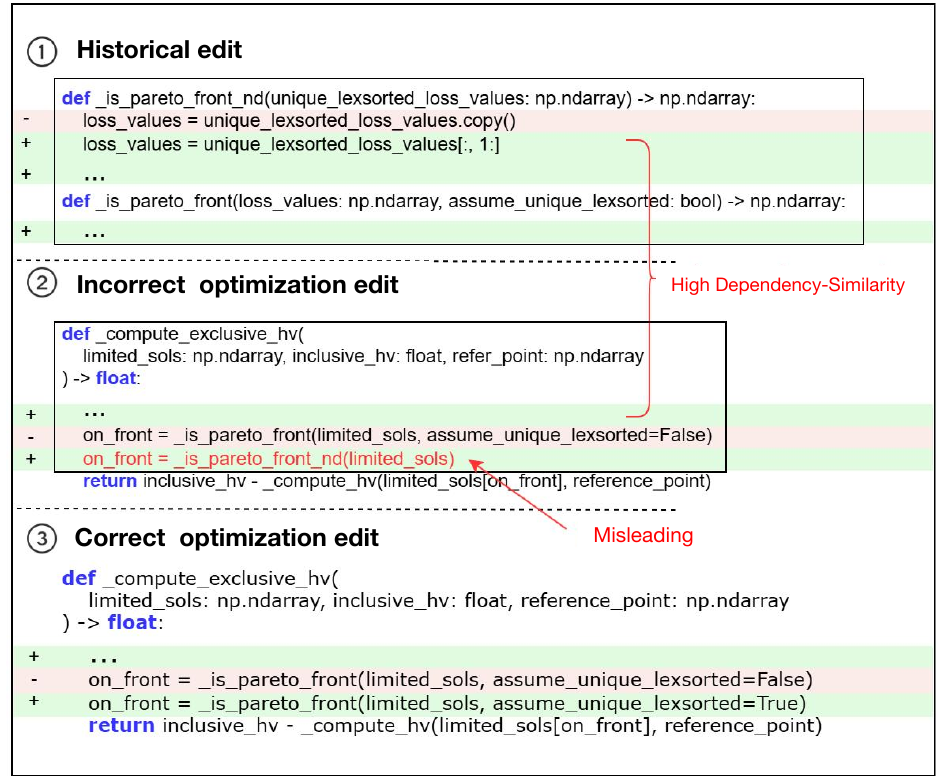}\vspace{-2mm}
     \caption{Motivating Example of Valid Associated Edits}
    \label{fig:case1}

\end{figure}

 \begin{figure}[ht]
    \centering
    \includegraphics[width=0.7\linewidth]{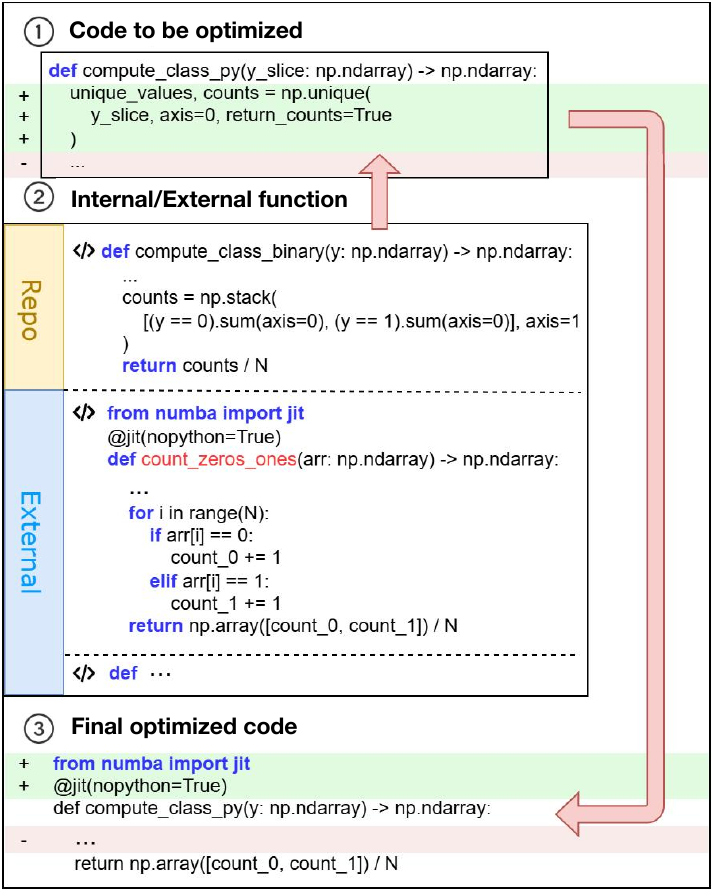}
     \caption{Motivating Example of Optimization Knowledge}
    \label{fig:case2}\vspace{-5mm}
\end{figure}

We use real-world examples to illustrate the two challenges mentioned in Sec~\ref{sec:intro} and motivate our work.
\begin{itemize}[leftmargin=*]
    \item \textbf{C1: Interference from invalid association editing.} 
    In software projects, historical edits offer valuable guidance for future changes~\cite{gupta2023grace, liu2024coedpilot}. However, using all historical edits is ineffective and impractical due to model input limits and noise. CoEdPilot~\cite{liu2024coedpilot} addresses this with a dependency score, but often struggles to filter out irrelevant/invalid edits, leading to incorrect or unnecessary modifications.
    As shown in Fig.\ref{fig:case1}, the dependency-based method incorrectly identifies both \texttt{\_is\_pareto\_front} and \texttt{\_is\_pareto\_front\_nd} as associated edits of the target function \texttt{\_compute\_exclusive\_hv}, due to their high dependency similarity. 
    If these misleading associations are incorporated into the editing process, the next edit may mistakenly select \texttt{\_is\_pareto\_front\_nd} (as shown in Fig.\ref{fig:case1} \ding{173}) rather than the correct \texttt{\_is\_pareto\_front} (as shown in Fig.\ref{fig:case1} \ding{174}), leading to test failures and functional degradation.
    
    \item \textbf{C2: Limited optimization knowledge.}
    Existing project-level code intelligence tasks focus on internal code reuse and context within a project, which is effective for code generation but insufficient for optimization. Projects may repeatedly use suboptimal patterns that mislead optimization efforts. In contrast, high-performance functions from external projects can offer valuable insights for improving code quality.
    As shown in Fig.~\ref{fig:case2}, the external function \texttt{count\_zeros\_ones} outperforms the internal function \texttt{compute\_class\_py} in terms of execution efficiency. Without incorporating external optimization knowledge, the final optimized function may result in suboptimal efficiency optimization. 
\end{itemize}

Therefore, to address \textbf{C1}, it is essential to accurately identify valid associated edits that provide meaningful guidance for efficiency optimization, while effectively filtering out invalid or misleading candidates. To achieve this, we propose a method that combines dependency score analysis with semantic similarity assessment (see Phase II of \tool). By jointly considering structural dependencies and semantic relevance, this approach significantly reduces the risk of introducing invalid edits and enhances the reliability of the optimization process.
For \textbf{C2}, augmenting with internal and external optimization knowledge is crucial. This includes leveraging external high-performance implementations. To this end, we fine-tune an \majorR{efficiency} optimizer capable of generating diverse and efficient solutions. The optimizer integrates both internal project knowledge and external optimization resources, enabling more informed decision-making and achieving comprehensive, effective project-level code \majorR{efficiency} optimization (see Phase III of \tool).

\section{Approach}

\begin{figure*}[!t]
    \centering
    \includegraphics[width=0.9\linewidth]{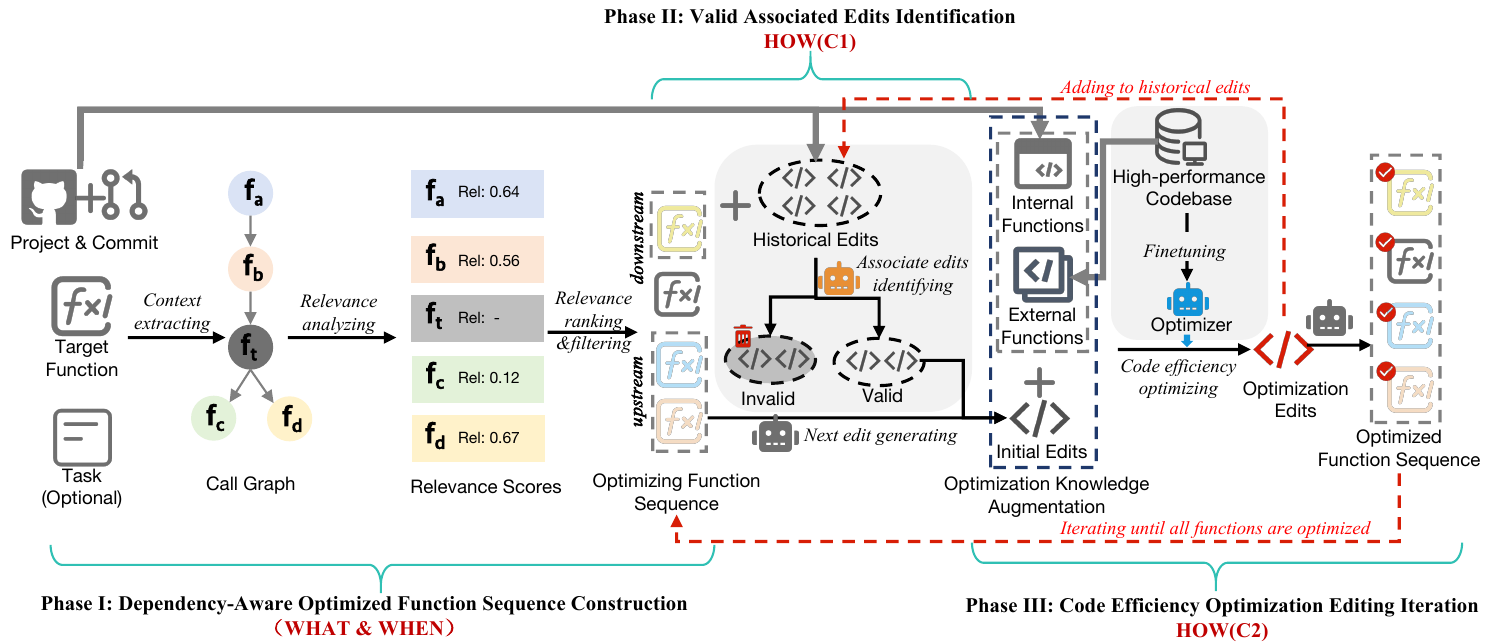}\vspace{-1mm}
    \caption{The Overall Framework of \tool}
    \label{fig:framework}
\end{figure*}

Fig.~\ref{fig:framework} shows the overall framework of \tool, which aims to optimize \majorR{the efficiency of} a target function and ensure the correctness and integrity of the overall project at the same time. 
To achieve this goal, \tool first analyzes code contexts to construct an optimizing function sequence for editing, and then identifies valid associated edits.
After that, it leverages valid associate edits along with both internal and external high-performance functions to iteratively optimize the functions in the optimizing function sequence.
Specifically, \tool contains three main phases: Dependency-Aware Optimizing Function Sequence Construction (Phase I), Valid Associated Edits Identification (Phase II), and \majorR{Code Efficiency Optimization} Editing Iteration (Phase III).

\subsection{Dependency-Aware Optimizing Function Sequence Construction}
A target function may depend on multiple related functions, requiring coordinated edits across them. In Phase I, \tool builds a sequence of functions to optimize, starting from the target function, through three key steps.

\subsubsection{Context Extraction}
We first extract the optimization context by constructing a call graph of the target function to identify candidate function sequences. The call graph includes all callee functions directly/indirectly called by the target function and caller functions that directly/indirectly call the target function. Specifically, we parse code into an abstract syntax tree (AST) using tree-sitter~\cite{treesitter}. Starting from the target function, we analyze function call nodes in the AST to identify its callees and callers, iterating over all functions to construct a complete call graph.
For example, as Fig.~\ref{fig:framework} shows, the caller functions of target function $f_t$ are $f_a$ and $f_b$, and the callee functions include 
$f_c$ and $f_d$.

\subsubsection{Relevance Analysis}\label{sec:rel}
After obtaining the call graph of the target function, we further measure the relevance between the target function and its caller and callee functions.
Following CoEdPilot~\cite{liu2024coedpilot}, we employ a dual approach considering both structural and semantic similarities. Structural similarity captures dependencies of functions, while semantic similarity ensures functional relevance by filtering out syntactically related but semantically irrelevant functions.
For dependency similarity, we utilize a transformer model (RobertaModel) in CoEdPilot~\cite{liu2024coedpilot} to estimate dependency scores between functions.  
For semantic similarity, we embed functions using the CodeBERT model and compute the cosine similarity.
By combining dependency similarity with semantic similarity, we can obtain relevance scores between the target function and its caller and callee functions.

\subsubsection{Relevance Filtering \& Ranking}

After computing the relevance scores between the target function and its caller and callee functions, we filter and rank these functions based on their scores to construct an optimizing function sequence, which serves as the edit order for project-level efficiency optimization.
Specifically, we first remove the caller and callee functions with relevance scores below a defined threshold (i.e., 0.5, referred to CoEdPilot~\cite{liu2024coedpilot}), as they do not demonstrate sufficient structural and semantic relevance to the target function. For example, the callee function 
$f_c$ is filtered out, as it gets a relevant score of 0.12.
Then, we rank the remaining relevant functions in descending order based on their relevance scores.
Note that following the bottom-up software development strategy~\cite{hollberg2019top}, the optimized order is determined as follows: callee functions (e.g., $f_d$) are ranked before the target function (e.g., $f_t$), while caller functions (e.g, $f_a$ and $f_b$) are ranked after the target function.
For example, the final optimizing function sequence for editing in Fig.~\ref{fig:framework} is $(f_d, f_t, f_a, f_b)$.

\subsection{Valid Associated Edits Identification }

As mentioned before, to deal with \textbf{C1}, we need to identify valid associated edits from the historical edits for each function to be optimized, avoiding misleading.
The historical edits are collected from the previous commits of the corresponding GitHub projects.
Before identifying, we first conduct a relevance analysis, following the approach in Sec~\ref{sec:rel}, to 
measure the relevance between the function to be optimized and each historical edit.
We can sort the historical edits in descending order of the relevance scores, which can efficiently filter out irrelevant edits that share no meaningful connection with the optimized function. 
\begin{figure}[h]
\vspace{-2mm}
    \centering
    \includegraphics[width=\linewidth]{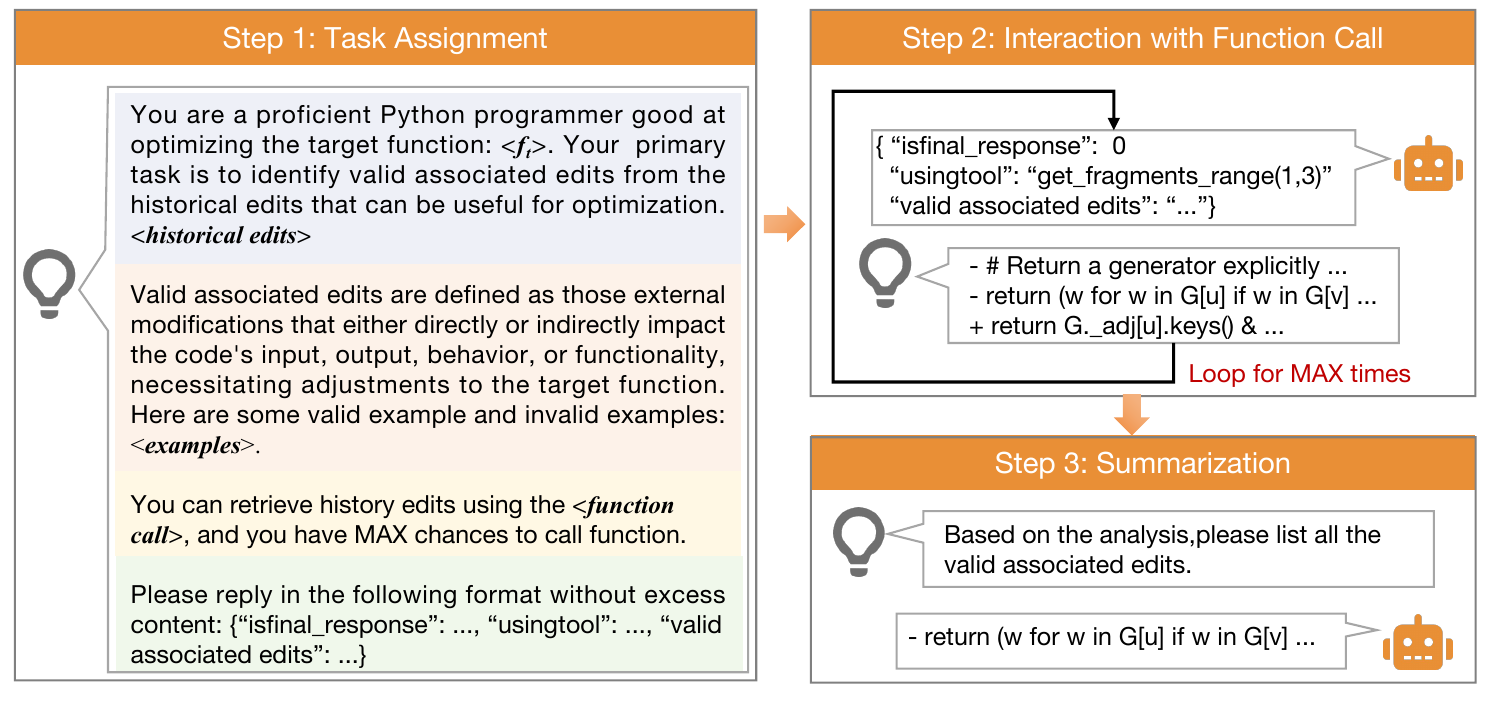}
    \caption{Agent Pipeline for Identifying Valid Associate Edits}
    \label{fig:agent}

\end{figure}

Then, we design an LLM-based agent to identify valid associated edits from the ranked historical edits.
Fig.~\ref{fig:agent} illustrates the pipeline of our agent, which contains three steps:
\begin{itemize}[leftmargin=*]
    \item \textbf{Step 1: Task assignment.} In this step, we design a prompt with information from four aspects. The first aspect is the system prompt, which asks LLM to identify associated edits from our given ranked historical edits. Second, we prompt LLM with the definition of valid associated edits, which can help LLM develop a precise understanding. Third, we provide function calls for the LLM to use. At last, we restrict the output format.
    \item \textbf{Step 2: Interaction with function call.} We design a function call (i.e., \texttt{get\_fragments\_range(i,j)}) for LLM to interact with. So that LLM can prioritize retrieving historical edits with higher relevance and iterate sequentially. The iteration continues until the LLM determines that sufficient valid associated edits have been obtained or reaches a maximum number of iterations (i.e., 10).
    \item \textbf{Step 3: Summarization.} We instruct the LLM to summarize and list all identified valid associated edits by integrating inputs from Step 1 and the information retrieved in Step 2. 
\end{itemize}

\subsection{\majorR{Code Efficiency Optimization Editing Iteration}}

\subsubsection{\majorR{Efficiency Optimizer Fine-tuning}}
To enhance the code efficiency optimization capability of general LLMs, we fine-tune an efficiency optimizer based on a small language model.

Before fine-tuning, we first construct a dataset specific to efficiency optimization tasks, which includes both function-level and project-level data. 
At the function level, we select optimization code pairs from the PIE dataset~\cite{shypula2023learning}, which is constructed from performance-improving edits made by human programmers across various competitive programming tasks in CodeNet~\cite{Puri2021Codenet}.
By filtering cases with performance improvements greater than 10\% to ensure that only meaningful optimizations are included, we collected a total of 3,450 function-level cases.
For the project level, we select 4,043 additional cases from GitHub, which are gathered through the benchmark construction process. These samples are performance-optimization related; however, due to limitations in execution feasibility, they are not included in the PEACExec.
This fine-tuning dataset is available in our replication package.

\begin{figure}[ht]
    \centering\vspace{-3mm}
    \includegraphics[width=0.8\linewidth]{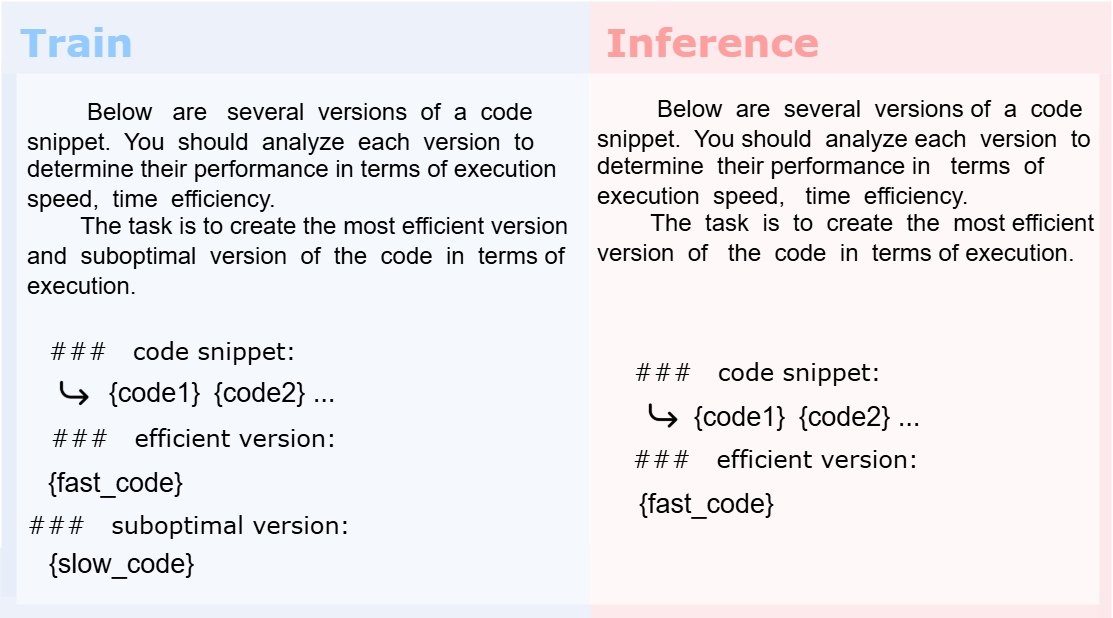}
    \caption{Prompts for Fine-tuning \majorR{Code Efficiency} Optimizer}\vspace{-3mm}
    \label{fig:optimizer}
\end{figure}

After that, we employ LoRA~\cite{Hu2021Lora} (Low-Rank Adaptation) to efficiently adapt the small language model to efficiency optimization tasks. 
During training, the model is presented with the original function alongside a set of semantically similar alternative implementations, as shown in Fig.~\ref{fig:optimizer} (left).
Each implementation in the training set is accompanied by a performance benchmark to guide the model in identifying the best-performing version from a set of similar code.
During inference, the model is instructed to generate the most efficient version of the function in terms of \majorR{efficiency} Fig.~\ref{fig:optimizer} (right).

This \majorR{efficiency} optimizer is then integrated into our \tool to assist the LLM to generate the most efficient code given a set of candidate codes. 
Details are described in section~\ref{sec:optimizing}

\subsubsection{\majorR{Code Efficiency} Optimizing}\label{sec:optimizing}

As previously mentioned, both internal and external \majorR{code efficiency} optimization knowledge is essential for project-level \majorR{efficiency} optimization. To generate the most efficient edits for a project without compromising its correctness and integrity, we adopt the following process for optimization.

Specifically, for each function (e.g., $f_d$) in the optimizing function sequence (constructed in Phase I), (1) we first feed the function, along with its valid associated edits (identified in Phase II), into the LLM to generate an initial edit for optimization. 
(2) Next, we use the code functionality embedding model (i.e., TransformCode~\cite{Xian2024transformcode}) to retrieve functionally similar code snippets from both internal and external libraries.
Here, internal libraries refer to code within the project itself, while external libraries consist of high-performance functions sourced from online repositories (i.e., Leetcode \cite{leetcode}, Numba \cite{numba}), which are also available in our released package.
(3) Then, the initial edit, along with the retrieved internal and external code snippets, is used to prompt our fine-tuned \majorR{code efficiency} optimizer, which generates an optimization edit. 
(4) At last, this optimization edit is integrated into the function by the LLM. This step is crucial, as the optimizer primarily focuses on \majorR{efficiency} improvements and may overlook correctness with respect to the broader project context. In this step, the LLM is prompted to incorporate the suggested edits while preserving contextual correctness. The detailed prompts can be found in our repository~\cite{ourpackage}. At the same time, the optimization edit is incorporated into the historical edits for the optimization of the next functions (e.g., $f_t$).

The iterative process continues until all functions in the optimizing function sequence have been optimized.

\subsection{Implement}
Our \tool is a hybrid framework that integrates an LLM (i.e., Llama-3.1-405B) with a fine-tuned \majorR{code efficiency} optimizer based on Llama3-8B.
We employ LoRA to fine-tune the optimizer, using a learning rate of $1 \times 10^{-5}$ and a batch size of 16. The model is trained for 4 epochs to ensure sufficient exposure to optimization examples while avoiding overfitting. Additionally, the LoRA rank is set to 32.

\section{\benchmark Construction}
As discussed earlier, previous studies on code efficiency optimization often primarily focus on function-level tasks, and existing datasets on project-level tasks~\cite{zhang2023repocoder, jimenez2023swe, liu2024coedpilot, Garg2022deepdev, garg2023rapgen} do not specifically address performance optimization or not available. 
To bridge this gap, we construct a project-level code optimization dataset, named \benchmark, from scratch. 

\subsection{\benchmark Overview}\label{sec:benchmarkOverview}
To assess the effectiveness of project-level code efficiency optimization techniques, our \benchmark is constructed based on GitHub issues related to \majorR{efficiency} optimization.
Each project-level optimization task in our \benchmark involves the following components:
\begin{itemize}[leftmargin=*]
    \item \textbf{Target function:} It is treated as the entry function to be optimized. It does not mean that the task is reduced to a function-level optimization; rather, full contextual information is available, and it may be necessary to consider cross-function and cross-file edits to optimize the target function effectively.
    \item \textbf{Task prompt (optional):} It refers to the commit message of the \majorR{efficiency} optimization issues. It is optional because some commits do not explicitly describe the optimization intent. If no task prompt is available, our framework uses a generic optimization instruction instead.
    \item \textbf{Historical commit:} It is used to collect historical edits of the optimization task. As illustrated by previous work~\cite{gupta2023grace, liu2024coedpilot}, historical edits play important roles in predicting the next edits.
    \item \textbf{Executable project:} It is the corresponding project for optimization, along with its docker-based runtime environment.
    \item \textbf{Test case:} It is collected from the corresponding optimization issue commits and used for evaluations. 
\end{itemize}

In total, our \benchmark contains 146 code \majorR{efficiency} optimization tasks covering 47 diverse projects, ensuring a comprehensive and representative assessment of project-level code \majorR{efficiency} optimization. 
Furthermore, we categorize the tasks in \benchmark into the following types:
\begin{itemize}[leftmargin=*]
    \item \textbf{Single-Function Project-Level Optimization:} It refers to simple optimization tasks that require editing only the target function itself to achieve project-level \majorR{code efficiency} optimization. This type consists of 80 tasks in our \benchmark.

    \item \textbf{Multi-Function Project-Level Optimization:} It refers to complex optimization tasks that require jointly editing multiple interdependent functions for project-level \majorR{code efficiency} optimization, while maintaining the consistency and functionality of projects. It covers 66 tasks in our \benchmark. 
\end{itemize}

\subsection{\benchmark Construction}\label{sec:Benchmark_Construction}
Here is the pipeline of \benchmark construction, including project selection, optimization task filtering, project environment setup \& test case validation.
\subsubsection{Project Selection}
Configuring executable environments for large-scale projects is challenging due to complex environment dependencies and unclear installation guidelines. 
To ensure that our benchmark closely aligns with real-world development scenarios, we carefully construct a dataset comprising over 2,000 of the top-starred Python projects from GitHub. 
Their varied coding styles, broad application domains, rich development patterns, and active maintenance contribute to a practical and scalable foundation for evaluating the effectiveness of techniques for project-level code \majorR{efficiency} optimization.

\subsubsection{Candidate Optimization Task Filtering}
Projects that have been maintained for an extended period often accumulate many issues and code commits over time.
Among the selected projects, the most active project contains more than 200,000 commits.
Considering the feasibility and effectiveness of projects, we first limited the candidate optimization tasks to the most recent 2,000 commits of each project, ultimately collecting approximately 440,000 candidate commits.
After that, we continuously selected commits relevant to \majorR{efficiency} optimization, with the following steps:
\begin{itemize}[leftmargin=*]
    \item \textbf{Step 1: Coarse-grained filtering:} \majorR{We first conduct a coarse-grained filtering by leveraging TF-IDF~\cite{qaiser2018text} to discover predefined \majorR{efficiency}-related keywords (e.g., ``optimize'', ``latency'', ``efficiency'' and ``fast''), which are available in our released package~\cite{ourpackage}) in commits' messages.} We filter out commits that do not contain such keywords, and reduce the number of candidates to around 16,000.
    \item \textbf{Step 2: Fine-grained filtering:} We then filter out commits that modify too many or too few lines/files from the 16,000 candidate commits.
    Specifically, we exclude commits that modify fewer than five lines, as these are considered trivial changes. Additionally, commits that modify more than 150 lines or involve five or more files are also removed, as they are deemed overly complex or redundant. Hence, 10,998 candidate commits remain.
    \item \textbf{Step 3: Semantic confirmation:} \majorR{To simplify our task, we only focus on each commit having a single optimization goal, i.e., improving execution speed. The project-level code optimization is quite a difficult task in terms of the complexity of the projects themselves, so to make the optimization task more focused, we only cover the task of execution speed optimization in the dataset. This simplification of the optimization objective does not affect the practicality of our method, since we notice in the process of collecting dataset that the number of execution speed commits is much more than that of memory optimization commits.}
    Specifically, for a more precise selection of performance-relevant commits, we employ the LLMs to make semantic confirmation of the remaining candidate commits. 
    We utilize commit messages along with their corresponding code diffs to prompt LLMs to rank each commit based on its relevance to \majorR{code efficiency}. 
    After this step, we retain  509 commits that show a strong correlation with \majorR{code efficiency} optimization issues, and such issues are candidate optimization tasks.
\end{itemize}

\subsubsection{Project Environment Setup \& Test Case Validation}
After obtaining high-quality commits and corresponding candidate optimization tasks (i.e., issues) related to project-level \majorR{code efficiency} optimization, configuring executable environments and running test cases for evaluation presents a continued challenge.
To ensure the reliability and feasibility of the selected tasks, we implement additional measures to validate that the test cases in commits are runnable and suitable for evaluating performance, along with ensuring the correctness of their respective environments.
\begin{itemize}[leftmargin=*]
    \item \textbf{Executable environment setup:} We create docker-based environments for each project to ensure proper setup of dependencies and configurations, enabling reliable \majorR{code efficiency} optimization.
    \item \textbf{Test case validation:} For each commit, we extract and prioritize test cases related to the function being optimized from commit patches. These test cases serve as essential evaluation metrics, ensuring that the \majorR{code efficiency} optimizations are verified under controlled conditions. 
\end{itemize}

To further enhance the quality of our benchmark, we manually double-check to ensure the collected commits are related to \majorR{efficiency} optimization, and filter out cases lacking sufficient test case coverage. 
This rigorous selection process results in a total of 146 commits across 47 diverse projects. 
Then, we collect all components needed (i.e., target function, optional task prompt, historical commits, executable projects, and test cases), and construct our \benchmark for project-level code \majorR{efficiency} optimization.
Note that the target function is manually selected based on the commit message and modification content.

\section{Experiment Setup}

\subsection{Research questions}
To systematically validate the effectiveness of \tool, we propose the following four research questions (RQs):
\begin{itemize}[leftmargin=*]
    \item \textbf{RQ1: What is the overall performance of our \tool in dealing with project-level code \majorR{efficiency} optimization?} To answer RQ1, we systematically compare \tool against popular baselines on \benchmark, aiming to investigate the capability of project-level code \majorR{efficiency} optimization. 
    \item \textbf{RQ2: How does \tool compare to baselines in terms of different optimization tasks?} For RQ2, we conduct a detailed comparative analysis of correctness and efficiency for \tool and baselines across single-function and multi-function optimization tasks.  
    \item \textbf{RQ3: What are the individual contributions of components designed in our \tool, and how do they influence the overall performance?} For RQ3, we conduct an ablation study by replacing different components of \tool with alternative implementations. This allows us to quantify the contribution of each component while also examining other factors that may influence performance in project-level code \majorR{efficiency} optimization tasks. 
    \item \textbf{RQ4: To what extent can the fine-tuned \majorR{code efficiency} optimizer optimize the overall performance of \tool? What is the significance and necessity of designing a hybrid framework?} For RQ4, we further investigate the capability of our fine-tuned performance optimizer by comparing it with popular proprietary LLMs. By replacing the optimizer with other models, we also investigate the design rationality of our hybrid framework.
\end{itemize}


\subsection{Baselines}\label{sec:baseline}
For overall performance evaluation, we select the following \majorR{six} relevant techniques as baselines, which cover both code editing techniques and \majorR{efficiency} optimization techniques. Due to the limited availability of research focusing on project-level optimization, we select both project-level and function-level optimization techniques as follows for evaluation.  
\begin{itemize}[leftmargin=*]
    \item \textbf{Instruction-Prompting}  ~\cite{Mishra2022reframing}: We directly prompt the LLM to improve the performance of the given target function, GPT-4o is used as the inference model.
    \item \textbf{Fine-Tuning}~\cite{shypula2023learning}: It refers to a method that enhances pre-trained LLMs for code optimization by fine-tuning on performance-annotated datasets. We adopt the performance-conditioned generation strategy proposed by Shypula et al.~\cite {shypula2023learning}.
    \item \textbf{SBLLM}~\cite{gao2024search}: It is a search-based framework that iteratively refines LLM-generated code optimizations. Specifically, in Stage 1 of SBLLM, we set $ns=3$, retaining all three generated candidates for downstream processing and ultimately selecting the best-performing one. In Stage 2, we utilize the SBLLM-provided knowledge base for retrieval. In Stage 3, GPT-4o is employed as the inference model.
    \item \textbf{DeepDev-PERF}~\cite{Garg2022deepdev}: \majorR{DeepDev-PERF is a deep learning-based method for software performance optimization proposed by Garg et al.~\cite{Garg2022deepdev}. Since the model and training data are not publicly released, we follow the training method described in their paper and use the dataset constructed by us in Section III.C of this paper to train a new model for comparison. Specifically, we extract statements, class attributes, caller-callee relationships, and signatures from the training data, and use these features to fine-tune a BART-large model for code generation.}
    \item \majorR{\textbf{RAPGen~\cite{garg2023rapgen}}: RAPGen is a method that fixes performance bugs in software by first retrieving an instruction from a knowledge-base of previous fixes. Its core technology uses this retrieved prompt to guide an LLM in a zero-shot setting to generate an effective code change.
    Since RAPGen does not release both source code and the retriever model, we reproduce the RAPGen according to their paper with the dataset in our dataset in Section III.C.}

    \item \textbf{CoEdPilot}~\cite{liu2024coedpilot}: It is a project-level code editing framework that integrates edit localization and feedback refinement, enabling context-aware code modifications based on prior edits. We adopt the open-source model provided by CoEdPilot~\cite{liu2024coedpilot}, with the optimization task specified via prompts.

\end{itemize}

To conduct an ablation study for investigating contributions of different components of \tool, we design the following variants by replacing different components of \tool with alternative implementations.
\begin{itemize}[leftmargin=*]
    \item \textbf{\tool\_w/o\_OFS:} It is designed to evaluate the performance of the optimized function sequence (OFS) constructed in phase I of \tool. Specifically, we ablate the optimized function sequence from two aspects for evaluation: \ding{172} using the target function as the only optimized function without considering other relevant functions; \ding{173} considering multiple relevant functions but with a random sequence for efficiency optimization. 
    \item \textbf{\tool\_w/o\_VAE:} It is designed to investigate the contribution of valid associated edits (VAE), which are identified in phase II of \tool.
    Specifically, we ablate the valid associated edits with two alternatives: \ding{172} no associated edits are given as an optimization assistant; \ding{173} associated edits extracted by the dependency analyzer~\cite{Jin2023evaluating} are given as an assistant of optimization.
    \item \textbf{\tool\_w/o\_OKA:} It is designed to investigate the contribution of optimization knowledge augmentation (OKA) in phase III.
    Specifically, we consider three settings for this variant: \ding{172} no function augmentation and only use the initial generated edits for efficiency optimization editing; \ding{173} use external functions to augment initial edits for efficiency optimization editing; \ding{174} use internal functions to augment initial edits for efficiency optimization editing.
\end{itemize}

Furthermore, in RQ4, we also replace our fine-tuned \majorR{code efficiency} optimizer in phase III with some popular LLMs to verify the performance of our fine-tuned optimizer and the necessity of this hybrid framework design.
Specifically, the alternatives include open-source and closed-source LLMs with large parameters (i.e., Llama-3.1-405B~\cite{llama3.1}, DeepSeek-V3-671B~\cite{deepseekv3}, GPT-4o~\cite{achiam2023gpt}, and Claude3.5~\cite{claude}), and one fine-tuned small LLM (i.e., CodeLlama-13B~\cite{codellama}).

\subsection{Metrics}
We employ the following metrics to assess the correctness and efficiency of the optimized functions, respectively:
\begin{itemize}[leftmargin=*]
    \item \textbf{Pass@1:} It represents the proportion of solutions that the model generates that correctly pass the test cases on the first try, which serves as an indicator of the correctness of the code editing~\cite{Liu2023is}.
    \item \textbf{Opt rate:} Prior research~\cite{Peng2025coffe} demonstrates that CPU instruction count is a more stable metric than execution time for evaluating code efficiency. Consequently, we use the opt rate to quantify the relative change in CPU instruction consumption before and after optimization. Since project-level code optimization may introduce test case inconsistencies before and after modification, we define opt rate as the instruction count ratio relative to the baseline (i.e.,  instruction-prompting with GPT-4o), capturing the performance impact of different optimization strategies. Specifically, it is can be computed as $Opt\ Rate=\frac{I_{\mathrm{instruction}}-I_{\mathrm{method}}}{I_{\mathrm{instruction}}}$, where $I_{instruction}$ represents the CPU instructions executed by the Instruction-Prompting baseline, $I_{method}$ denotes the CPU instruction count of the evaluated optimization method.
    \item \majorR{\textbf{Speedup:} The ratio $\frac{gt}{o}$ of CPU instruction count of ground truth solution (i.e., human-written patch) $gt$ to the CPU instruction count of a code solution $o$. A speedup value exceeding 1.0 signifies superior performance of the generated code compared to the human benchmark. Conversely, a value approaching 1.0 from above indicates that the automated optimization's performance is converging on human-level quality. The primary objective is to maximize this value, striving for generated code that meets or exceeds human performance.}
\end{itemize}

\section{Evaluation}
\subsection{Overall Performance of \tool (RQ1)}

\begin{table}[]
\centering
\caption{Overall Performance of \tool in Project-level Code Efficiency Optimization}\label{tab:overall_performance}
\renewcommand{\arraystretch}{0.8} 
\majorR{\begin{tabularx}{1\columnwidth}{c|A|A|A} 
\toprule
\multirow{2}{*}{\textbf{Method}}  & \multicolumn{3}{c|}{\textbf{\benchmark}} \\
\cline{2-4}  
& \textbf{Pass@1} & \textbf{Opt Rate} & \textbf{Speedup} \\  
\midrule
Instruction-Prompting & 52.7 & - & 0.446\\
Fine-Tuning &  48.6 & +24.5 & 0.591\\
DeepDev-PERF &  34.9 & +11.6  & 0.505\\
RAPGen & 41.1 & +12.2& 0.508\\
SBLLM &  58.2 & +28.5 & 0.624\\
CoEdPilot &  45.9 & +7.8  & 0.484\\
\midrule
\tool & \textbf{69.2} & \textbf{+46.9} & \textbf{0.840}\\ 
\bottomrule
\end{tabularx}}  
\end{table}

 For RQ1, we conduct experiments on \benchmark to show the overall performance of \tool in project-level code efficiency optimization compared to the other baselines.
 \majorR{We select six baselines, including three important code optimization methods (i.e.,  SBLLM~\cite{gao2024search}, Instruction-Prompting with GPT-4o and Fine-Tuning with CodeLlama proposed by Shypula et al.~\cite{shypula2023learning}), three code performance optimization methods (i.e., Deepdev-PERF~\cite{Garg2022deepdev}, RAPGen~\cite{garg2023rapgen}, SBLLM~\cite{gao2024search}), and the SOTA code editing method (i.e, CoEdPilot~\cite{liu2024coedpilot}). 
 We evaluate the performance from both correctness and efficiency by measuring pass@1, opt rate \& speedup, respectively.} Note that (1) Test
cases that do not pass the correctness checks are excluded from
opt rate \& speedup computation. (2) We consider Instruction-Prompting as the performance standard for measuring the opt rate of each method, as it is the most direct method leveraging the capabilities of GPT-4o, the most popular and SOTA LLM. Table.~\ref{tab:overall_performance} shows the overall performance of our \tool and baselines in project-level code optimization.

 \subsubsection{Correctness}
 As shown in Table~\ref{tab:overall_performance}, our \tool achieves a pass@1 correctness rate of 69.2\%, significantly outperforming all baselines by 11.0\% to 23.3\%. This demonstrates that \tool can generate accurate, reliable optimization edits without compromising correctness or project quality.

SBLLM delivers the second-best pass@1 score (58.2\%) due to its broader search space, enabling exploration of diverse optimization candidates. However, its lack of fine-grained function dependency modeling limits its correctness. For example, when optimizing \texttt{\_compute\_exclusive\_hv} (Fig.~\ref{fig:case1} \ding{173}), SBLLM introduces a pruning strategy that skips Pareto computation but fails to account for parent function dependencies, leading to incorrect results.
\majorR{DeepDev-PERF, RAPGen, and CoEdPilot perform the worst, with 37.8\%, 41.1\%, and 45.9\% correctness. Although DeepDev-PERF and RAPGen is designed for project-level code optimization, it struggles with complex function interactions.} Additionally, CoEdPilot often introduces invalid edits when conducting project-level code editing. As shown in Fig.~\ref{fig:case1}, it mistakenly replaces \texttt{\_is\_pareto\_front} with \texttt{\_is\_pareto\_front\_nd}, causing computational errors due to incorrect dependency-based retrieval.
Instruction-Prompting, using GPT-4o, achieves 52.7\%. Its prompt-only approach limits its ability to model cross-functional dependencies, leading to incomplete or incorrect edits. Fine-Tuning performs slightly worse (48.6\%), as its limited generalization hampers handling complex optimization tasks.


\subsubsection{Efficiency}
As shown in Table~\ref{tab:overall_performance}, our \tool achieves an impressive opt rate of +46.9\%, surpassing all baselines by 18.4\% to 39.1\%. This highlights \tool's ability to generate optimized code that significantly improves project-level runtime performance.
\majorR{Additionally, our \tool achieves 0.840 in speedup, which indicates that \tool's optimization performance is closest to that of human-written patches, outperforming all other baselines.}
The efficiency gains stem from \tool's integration of both internal and external optimization knowledge. For example, when optimizing \texttt{compute\_class\_py} (Fig.~\ref{fig:case2}), it retrieves \texttt{compute\_class\_binary} (internal) and \texttt{count\_zeros\_ones} (external, Numba JIT optimized), ultimately selecting the most efficient implementation.

In comparison, SBLLM achieves an opt rate of +28.5\% by exploring diverse optimization candidates, ranking the second. 
However, its limited modeling of function dependencies and execution contexts constrains further improvements.
DeepDev-PERF delivers +11.6\% opt rate and Fine-Tuning delivers +24.5\% opt rate, which shows that the lack of deeper analysis and optimization heuristics limits their capability in complex, repository-scale optimization scenarios. CoEdPilot shows minimal improvement (+7.8\%), as it lacks mechanisms to target performance bottlenecks effectively.


\begin{tcolorbox}[width=\linewidth-2pt,boxrule=0pt,top=2pt, bottom=2pt, left=2pt,right=2pt, colback=gray!20,colframe=gray!20]
\textbf{Answer to RQ1:} Our \tool outperforms all baseline methods, achieving a high correctness of 69.2\% pass@1 and delivering a substantial opt rate of +46.9\% and speedup of 0.840. It highlights \tool's effectiveness in improving both the correctness and efficiency of project-level code efficiency optimization.
\end{tcolorbox}

\subsection{Comparative Analysis (RQ2)}
\begin{figure}
    \centering
    \includegraphics[width=1\linewidth]{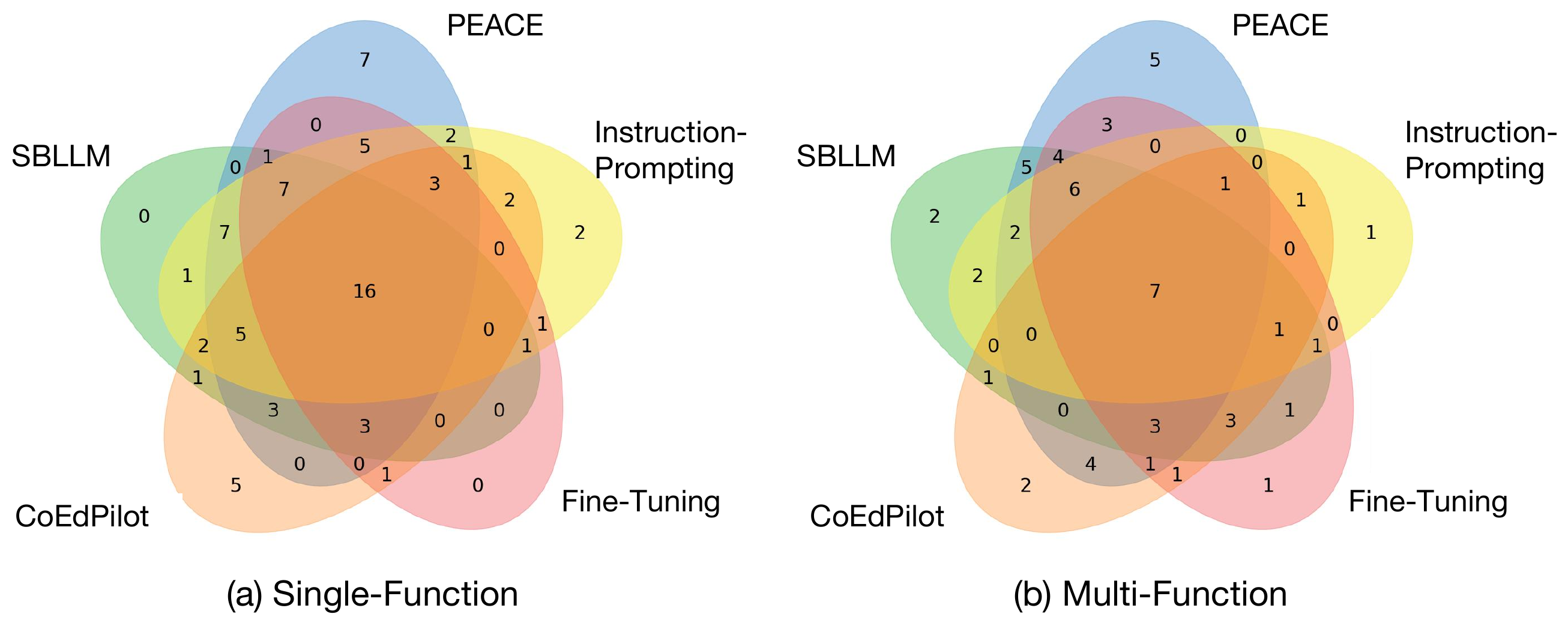}
    \caption{Correctness Overlap Between \tool and Baselines} \vspace{-5mm}
    \label{fig:correctness_overlap}
\end{figure}

\begin{figure}
    \centering 
    \includegraphics[width=1\linewidth]{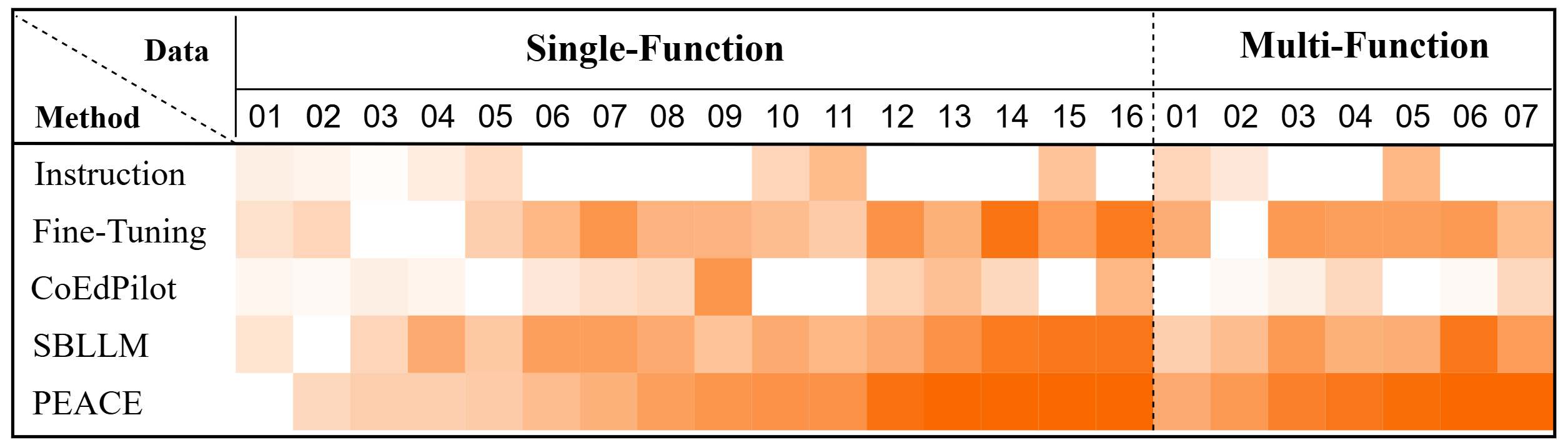}
    \caption{Efficiency Comparison Between \tool and Baselines on Correct Cases}\vspace{-3mm}
    \label{fig:efficiency}
    \vspace{-3mm}
\end{figure}

To further evaluate optimization performance, we conduct a comparative analysis from both correctness and efficiency aspects. 
\majorR{Note that, for clarity of presentation, we select the five methods with the highest optimization correctness for further comparison, including our \tool, SBLLM, Instruction-Prompting, Fine-Tuning, and CoEdPilot.}

\subsubsection{Correctness}
Figure~\ref{fig:correctness_overlap} compares the correctness of \tool and baseline methods in project-level optimization. The analysis covers two task types: single-function and multi-function, providing valuable insights into how \tool compares to the baselines.

For single-function tasks shown in Figure~\ref{fig:correctness_overlap} (a), \tool demonstrates a significant advantage over the baselines, successfully optimizing seven unique tasks. While there is notable overlap between \tool and SBLLM, it is clear that \tool outperforms SBLLM, particularly in optimizing certain single-function tasks where SBLLM falls short.
For multi-function tasks, as shown in Figure~\ref{fig:correctness_overlap} (b), \tool again shows a clear advantage over the baselines, excelling in optimizing complex tasks with high correctness across five unique tasks. 
While there is substantial overlap between \tool and CoEdPilot, indicating that CoEdPilot is a strong baseline for performing correctness edits at the project level, it does not outperform \tool in optimizing multi-function tasks.


\subsubsection{Efficiency}
For the efficiency comparison, we focus on the 23 optimization tasks that all methods successfully edit, including 16 single-function tasks and 7 multi-function tasks. Figure~\ref{fig:efficiency} uses color intensity to represent optimization performance, where darker shades indicate higher performance improvements.
For single-function tasks, \tool consistently outperforms all baseline methods. Figure~\ref{fig:efficiency} demonstrates that \tool achieves the highest Opt rates in several tasks, particularly when compared to Instruction-Prompting and CoEdPilot, which show more sporadic improvements. 
While baselines like SBLLM and Fine-Tuning show some improvement in specific tasks, their performance is generally less consistent and lower than that of \tool.
For multi-function tasks, \tool shows a significant advantage over baselines. Although CoEdPilot and SBLLM show some degree of improvement, their performance fails to match the high level achieved by \tool. 
This underscores \tool's ability to handle more complex optimization tasks effectively, where other methods struggle or exhibit less substantial improvements.

\begin{tcolorbox}[width=\linewidth-2pt,boxrule=0pt,top=2pt, bottom=2pt, left=2pt,right=2pt, colback=gray!20,colframe=gray!20]
\textbf{Answer to RQ2:} \tool outperforms baselines in both correctness and efficiency across single-function and multi-function optimization tasks, achieving better results, especially in complex tasks. 
\end{tcolorbox}

\subsection{Ablation Study (RQ3)}
To assess the impact of key components in \tool, we perform an ablation study by systematically removing or modifying them. We construct three variants: \tool\_w/o\_OFS, \tool\_w/o\_VAE, and \tool\_w/o\_OKA. Each variant includes two or three alternative versions of the removed component, marked as \ding{172}, \ding{173}, and \ding{174} (see Sec~\ref{sec:baseline} for details). We evaluate all variants on correctness (pass@1) and efficiency (opt rate \& speedup). Results are shown in Table~\ref{tab:ablation_study}.

\subsubsection{Ablation of OFS}
For optimizing function sequences (OFS), limiting edits to a single target function (\tool\_w/o\_OFS\_\ding{172}) slightly improves correctness (70.5\% vs. 69.2\%) due to a smaller editing scope, but significantly reduces the opt rate to +29.5\%, and the speedup drops to 0.633, limiting global performance gains. Allowing multi-function edits in random order (\tool\_w/o\_OFS\_\ding{173}) raises the opt rate to +42.7\% but drops correctness to 56.2\%, showing instability from uncontrolled edit order. These results highlight the need to coordinate function selection and edit order for effective optimization.

\subsubsection{Ablation of VAE}
Removing historical edit information (\tool\_w/o\_VAE\_\ding{172}) significantly reduces correctness (47.2\% pass@1), highlighting the importance of prior knowledge for consistent and accurate edits. Interestingly, the efficiency remains moderately high (+34.3\% of opt rate and 0.679 of speedup), indicating that some performance gains are possible, though often at the expense of reliability. Using a dependency-based retrieval method~\cite{Jin2023evaluating} (\tool\_w/o\_VAE\_\ding{173}) improves correctness to 59.6\% and opt rate to +41.9\%, but still falls short of \tool. This suggests that structural dependencies alone are insufficient; semantic context is also crucial for identifying valid associated edits that truly support effective optimization.

\subsubsection{Ablation of OKA}
For optimization knowledge augmentation, using only initial edits without any internal or external knowledge (\tool\_w/o\_OKA\_\ding{172}) maintains correctness (69.6\%) but significantly reduces the opt rate to +18.4\% and the speedup is only 0.547, showing that our augmentation of the optimizer mainly boosts efficiency.
Using only external high-performance functions (\tool\_w/o\_OKA\_\ding{173}) improves opt rate (+41.7\%) but slightly lowers correctness (68.9\%), indicating efficiency gains at the cost of some inconsistency.
In contrast, relying only on internal functions (\tool\_w/o\_OKA\_\ding{174}) achieves the highest correctness (71.9\%) but reduces the opt rate to +24.3\%, suggesting internal knowledge preserves consistency but limits optimization potential.

\begin{table}
\centering
\caption{The Result of Ablation Study}\label{tab:ablation_study}
\vspace{-2mm}
\majorR{\begin{tabular}{c|c|ccc}
\toprule
\multirow{2}{*}{\textbf{Method}}                & \multirow{2}{*}{\textbf{Alternative}} & \multicolumn{3}{c}{\textbf{Metric}}                     \\ \cline{3-5} 
                                                &                                       & \multicolumn{1}{c}{\textbf{Pass@1}} & \textbf{Opt Rate} & \textbf{Speedup} \\ \hline
\multirow{2}{*}{\tool\_w/o\_OFS} & \ding{172}                                     & \multicolumn{1}{c}{70.5}            & +29.5 &0.633            \\ \cline{2-5} 
                                                & \ding{173}                                       & \multicolumn{1}{c}{56.2}            & +42.7 &0.778           \\ \hline
\multirow{2}{*}{\tool\_w/o\_VAE} & \ding{172}                                       & \multicolumn{1}{c}{47.2}            & +34.3 &0.679           \\ \cline{2-5} 
                                                & \ding{173}                                       & \multicolumn{1}{c}{59.6}            & +41.9 & 0.786          \\ \hline
\multirow{3}{*}{\tool\_w/o\_OKA} & \ding{172}                                       & \multicolumn{1}{c}{69.6}            & +18.4 &0.547            \\ \cline{2-5} 
                                                & \ding{173}                                      & \multicolumn{1}{c}{68.9}            & +41.7 &0.765            \\ \cline{2-5} 
                                                & \ding{174}                                       & \multicolumn{1}{c}{71.9}            & +24.3 &0.589           \\ \hline
\tool                            & -                                     & \multicolumn{1}{c}{\textbf{69.2}}            & \textbf{+46.9} & \textbf{0.840}           \\ \bottomrule
\end{tabular}}
\end{table}

\begin{tcolorbox}[width=\linewidth-2pt,boxrule=0pt,top=2pt, bottom=2pt, left=2pt,right=2pt, colback=gray!20,colframe=gray!20]
\textbf{Answer to RQ3}: Our \tool is well-designed, and all components (i.e., optimized function sequences, efficient association edits, and optimized knowledge augmentation) can be integrated together to contribute to the superior performance of \tool in both correctness and efficiency.
\end{tcolorbox}

\subsection{Performance of Optimizer (RQ4)}
To assess the effectiveness of our hybrid optimization framework, which combines an LLM (i.e., Llama3.1-405B) with a fine-tuned small model (i.e., Llama3-8B), we conducted comprehensive experiments by replacing our fine-tuned performance optimizer with other models, including open-source and closed-source LLMs and another fine-tuned small model.

According to Table~\ref{tab:model_performance}, our \majorR{code efficiency optimizer} (i.e., fine-tuned with Llama-3-8B) achieves a pass@1 correctness of 69.2\% and delivers a +46.9\% opt rate and 0.840 in speedup, outperforming both larger fine-tuned models and closed-source LLMs in execution efficiency.
Table~\ref{tab:model_performance} also shows that while LLMs with large parameters, such as GPT-4o and DeepSeek-V3, achieve higher correctness (71.2\% and 70.8\% on Pass@1, respectively), they offer lower efficiency gains compared to our fine-tuned small models. 
Specifically, Llama-3.1-405B achieves only +32.9\% in opt rate and 0.665 in speedup, significantly lagging behind the +46.9\% achieved by our optimizer and far away from the efficiency of human-written patches.
It demonstrates the advantage of our hybrid framework, where LLMs ensure correctness and semantic consistency at the project level, while fine-tuned small models, optimized for performance, deliver superior execution efficiency. 
The integration of both model types allows our \tool to balance correctness and efficiency at the same time.

\begin{table}
\centering
\caption{Effectiveness of Our \majorR{Code Efficiency} Optimizer}\label{tab:model_performance}
\renewcommand{\arraystretch}{} 
\majorR{\begin{tabularx}{\columnwidth}{B|B|ccc}  
\toprule
\multirow{2}{*}{\textbf{Type}} & \multirow{2}{*}{\textbf{Model}} &\multicolumn{2}{c}{\textbf{Metric}} \\
\cline{3-5}
& & \textbf{Pass@1} & \textbf{Opt Rate} & \textbf{Speedup}\\ 
\midrule
\multirow{4}{*}{Closed-source} & Llama-3.1-405B & 66.7 & +32.9 & 0.665\\
\cline{2-5}
& DeepSeek-V3 & 70.8 & +40.5 & 0.750\\
\cline{2-5}
& GPT-4o & 71.2 & +37.2 & 0.710\\
\cline{2-5}
& Claude-3.5 & 69.8 & +33.6 & 0.672\\
\hline
\multirow{4}{*}{Fine-Tuned} & CodeLlama-13B & 65.8 & +46.4 & 0.832 \\
\cline{2-5}
& Llama-3-8B
(Our)& \textbf{69.2} & \textbf{+46.9}  & \textbf{0.840}\\
\bottomrule
\end{tabularx}}
\end{table}

\begin{tcolorbox}[width=\linewidth-2pt,boxrule=0pt,top=2pt, bottom=2pt, left=2pt,right=2pt, colback=gray!20,colframe=gray!20]
\textbf{Answer to RQ4:} 
Our fine-tuned performance optimizer outperforms other baselines. By combining a fine-tuned LLaMA-3-8B with Llama-3.1-405B, our hybrid framework \tool can leverage the small model's strength in performance optimization and the LLM's capability in code generation. This design provides an efficient and reliable solution for project-level code optimization, ensuring both high correctness and execution efficiency.
\end{tcolorbox}

\section{Discussion}
\majorR{

\subsection{Computational Costs and Hybrid Architecture Design}
\tool adopts a hybrid architecture instead of fine-tuning a large language model (LLM, Llama3-1405B) to balance correctness, efficiency, and cost. It combines two lightweight models, a pre-trained CodeBERT for relevance analysis in Phase I and a fine-tuned optimizer for edit refinement in Phase III, with two stages of LLM interaction for validation and optimization.
Direct fine-tuning of a large LLM would be prohibitively expensive and unsuitable for iterative optimization. \tool therefore relies on the LLM mainly for project-wide reasoning and dependency-aware editing, while a smaller fine-tuned model (Llama3-8B) handles performance-oriented optimization at lower cost. Both lightweight models run within seconds on an Nvidia A800, so the primary overhead comes from LLM interactions. This cost is controlled through relevance filtering in Phase I, which reduces token usage and model calls. Most tasks converge within five interaction rounds, and a hard cap of ten iterations prevents unnecessary overhead. These measures ensure that PEACE remains computationally practical while achieving improvements in correctness and efficiency.}

\subsection{Threads to Validity}
\subsubsection{Internal Validity}
Our work has three main limitations. First, our evaluation relies on correctness (pass@1) and execution efficiency (opt rate). While pass@1 assesses code correctness, our opt rate differs from typical optimization tasks, it compares performance against GPT-4o, since our GitHub-sourced project-level tasks lack pre-optimization test cases.
Second, the baselines include SBLLM (function-level), DeepDev-PERF (repo-level), CoEdPilot (code editing), and LLM-based methods (Instruction-Prompting, Fine-Tuning).
Third, we use static analysis to construct call graphs, which may miss dynamic calls (e.g., via inheritance or polymorphism). While effective for capturing explicit dependencies, future work will explore integrating dynamic analysis to improve accuracy.
\subsubsection{External Validity}
Our benchmark, \benchmark, is built from popular, highly starred Python projects on GitHub. This ensures it reflects widely used open-source projects but may limit generalization to other programming languages, such as C++ or Java, or to less-maintained codebases. Additionally, although our experiments include both open-source and closed-source, the rapid evolution of LLM architectures and fine-tuning techniques could impact future results. Therefore, our findings should be understood within the context of the current generation of models.

\section{Related Work}

\subsection{Code Optimization}

With the rise of large language models (LLMs), code optimization has advanced at both the function and project levels.
At the function level, PIE~\cite{shypula2023learning} benchmarks LLMs for performance-aware code generation using prompting. Later works~\cite{Shrivastava2023Repository, Gao2024Learning} improve results through prompting and fine-tuning. Techniques like STOP~\cite{zelikman2024self} and retrieval-based models~\cite{Brown2020language, Levine2022standing, Shi2023replug} enhance quality via feedback and external knowledge. SBLLM~\cite{gao2024search} adds search-based refinement, while others~\cite{ Xia2023Keep, Lu2022reacc} incorporate profiling and static analysis to guide optimization.
At the project level, methods like DeepDev-PERF~\cite{Garg2022deepdev} fine-tune models with project-wide context. RAPGen~\cite{garg2023rapgen} uses performance-related commits to retrieve optimization examples. CodeDPO~\cite{zhang2024codedpo} applies iterative self-validation, and COCOGEN~\cite{bi2024iterative} combines compiler feedback with context retrieval. 
Despite this progress, general-purpose, scalable solutions for project-level performance optimization remain limited, highlighting the need for further research.

\subsection{Code Editing}
Code editing has gained growing interest, with various methods developed to predict and generate code changes.
Early approaches like Codit~\cite{Chakraborty2022codit} used tree-based neural networks to model edits based on AST hierarchies. Recoder~\cite{Zhu2021a} improved this by combining AST and code readers to better understand code structure and boost prediction accuracy. Later methods explored pre-training and transformer models. CURE~\cite{Jiang2021cure} used pre-trained models for program repair, showing the benefits of large-scale pre-training for producing correct edits. CoditT5~\cite{Zhang2023coditt5} extended CodeT5~\cite{Wang2021codet5} by training on inputs that include comments and code changes, generating structured edit plans for more accurate edits. Recent work focuses on prompt-based learning with large language models. GRACE~\cite{gupta2023grace} fine-tunes LLMs with prompts that include related code updates, improving edit quality. CoEdPilot~\cite{liu2024coedpilot} also uses LLMs to identify relevant prior edits and predict where changes should be made, enabling more precise and context-aware editing.
Existing code editing methods focus on general edits, limiting their use in performance optimization. In contrast, we propose a hybrid framework tailored for targeted, effective code optimization.


\section{Conclusion}
We propose \tool, a hybrid framework for project-level performance optimization via automatic code editing. Unlike function-level approaches, \tool captures interdependencies among functions to optimize multiple components while preserving correctness.
Evaluated on \benchmark, including 146 tasks from 47 Python GitHub projects, our \tool significantly outperforms state-of-the-art methods like CoEdPilot, SBLLM, and DeepDev-PERF in both correctness and efficiency.
\tool offers a robust solution for real-world software optimization, with future work aiming to extend language support and improve integration with external libraries.

\section*{Acknowledgment}
This research/project was partially supported by the National Natural Science Foundation of China ( No.62302437, No. 62202420), the Fundamental Research Funds for the Central Universities 226-2025-00004, and Zhejiang Provincial Natural Science Foundation of China (No.LZ25F020003).


\bibliographystyle{IEEEtran}
\balance
\bibliography{sample-base}
\end{document}